# Binary interaction dominates the evolution of massive stars[1]

H. Sana[1]*, S.E. de Mink[2,3]†, A. de Koter[1,4], N. Langer[5], C.J. Evans[6], M. Gieles[7], E. Gosset[8], R.G. Izzard[5], J.-B. Le Bouquin[9], F.R.N. Schneider[5]

[1]Astronomical Institute 'Anton Pannekoek', Amsterdam University, Science Park 904, 1098 XH, Amsterdam, The Netherlands

[2]Space Telescope Science Institute, 3700 San Martin Drive, Baltimore, MD 21218 USA

[3]Department of Physics and Astronomy, Johns Hopkins University, Baltimore, MD 21218, USA

[4]Astronomical Institute, Utrecht University, Princetonplein 5, 3584 CC, Utrecht, The Netherlands

[5]Argelander-Institut für Astronomie, Universität Bonn, Auf dem Hügel 71, 53121 Bonn, Germany

[6]UK Astronomy Technology Centre, Royal Observatory Edinburgh, Blackford Hill, Edinburgh, EH9 3HJ, UK

[7]Institute of Astronomy, University of Cambridge, Madingley Road, Cambridge CB3 0HA, United Kingdom

[8]F.R.S.-FNRS, Institut d'Astrophysique, Liège University, Allée du 6 Août 17, B-4000 Liège, Belgium

[9]UJF-Grenoble 1 / CNRS-INSU, Institut de Planétologie et d'Astrophysique de Grenoble (IPAG) UMR 5274, Grenoble, France

*Correspondence to: H.Sana@uva.nl

†Hubble Fellow

**The presence of a nearby companion alters the evolution of massive stars in binary systems, leading to phenomena such as stellar mergers, X-ray binaries and gamma-ray bursts. Unambiguous constraints on the fraction of massive stars affected by binary interaction were lacking. We simultaneously measured all relevant binary characteristics in a sample of Galactic massive O stars and quantified the frequency and nature of binary interactions. Over seventy per cent of all massive stars will exchange mass with a companion, leading to a binary merger in one third of the cases. These numbers greatly exceed previous estimates and imply that binary interaction dominates the evolution of massive stars, with implications for populations of massive stars and their supernovae.**

---

With masses larger than 15 times that of our Sun (*1*), stars of spectral type O are rare (*2*) and short lived (*3*). Nevertheless, through their large luminosities, strong stellar winds and powerful explosions, massive stars heat and enrich surrounding gas clouds in which new generations of stars form (*4*) and drive the chemical evolution of galaxies (*5*). Massive stars end their lives in luminous explosions, as core-collapse supernovae (CCSNe) or gamma-ray bursts (GRBs), which can be observed throughout most of the Universe.

In a binary system, the evolutionary path of a massive star is drastically altered by the presence of a nearby companion (*6-8*). Because stars expand as they evolve, those in pairs with orbital periods up to about 1500 days exchange mass (*6*). The more massive star can be stripped of its entire envelope, and thus loses much of its original mass. The companion star gains mass and angular momentum, which trigger mixing processes in the stellar interior and modifies its evolutionary path (*3*). In very close binaries, the two stars may even merge. The nature of the binary interaction is largely determined by the initial orbital period and mass ratio. The relative roles of interaction scenarios and the overall importance of binary- versus single-star evolution so far remain uncertain because of the paucity of direct measurements of the intrinsic distributions of orbital parameters (*9-14*).

Here, we homogeneously analyze the O star population of six nearby Galactic open stellar clusters and simultaneously measure all the relevant intrinsic multiplicity properties (*15*). Our observational method, spectroscopy, is sensitive to orbital periods as long as 10 years (*13*), which corresponds to the relevant period range for binary interaction (*6*). In a spectroscopic binary the periodic Doppler shift of spectral lines allows the determination of the radial velocity, and hence of the orbital motion, of one (`single-lined' spectroscopic binary) or both (`double-lined' spectroscopic binary) stars. Given sufficient orbital-phase coverage, the orbital period (*P*), the eccentricity (*e*) and, for double-lined spectroscopic binaries, the mass-ratio (*q*) follow from Kepler's laws.

Our sample contains 71 single and multiple O-type objects (see supporting online text §A). With 40 identified spectroscopic binaries, the observed binary fraction in our sample is $f_{obs}$ = 40/71 = 0.56. We combined observations obtained with the *Ultraviolet and Visible Echelle Spectrograph* (UVES) at the *Very Large Telescope* for long-period systems with results from detailed studies of detected systems in the individual clusters (*16-21*). In total, 85% and 78% of our binary systems have, respectively, constrained orbital periods and mass-ratios. This allowed us to build statistically significant observed period and mass-ratio distributions for massive stars (Fig. 1), which are representative of the parameter distributions of the Galactic O star population (*13*).

The precise fraction of interacting O stars, and the relative importance of the different interaction scenarios is determined by the distributions of the orbital parameters. The observed distributions result from the intrinsic distributions and the observational biases (see supporting online text §B). To uncover the intrinsic distributions, we simulate observational biases using a Monte Carlo approach that incorporates the observational time series of each object in our sample. We adopt power laws for the probability density functions of orbital periods (in $\log_{10}$ space), mass-ratios and eccentricities with exponents π, κ and η, respectively (Table S3 and Fig. S3). These power-law exponents and the intrinsic binary fraction $f_{bin}$ were simultaneously determined by a comparison of simulated populations of stars with our sample allowing for the observational biases. We determined the accuracy of our method by applying it to synthetic data.

Compared to earlier attempts to measure intrinsic orbital properties (*9-14*): (*i*) the average number of epochs per object in our sample is larger by up to a factor of five, making binary detection more complete, (*ii*) over three quarters of our binaries have measured orbital properties, which allowed us to directly model the orbital parameter distributions, (*iii*) the orbital properties cover the full range of periods and mass-ratios relevant for binary interaction. We are thus better equipped to draw direct conclusions on the relative importance of various binary interaction scenarios.

We find an intrinsic binary fraction of $f_{bin}$ = 0.69 ± 0.09, a strong preference for close pairs ($\pi$ = -0.55 ± 0.2) and a uniform distribution of the mass ratio ($\kappa$ = -0.1 ± 0.6) for binaries with periods up to about nine years. Comparison of the intrinsic, simulated and observed cumulative distributions of the orbital parameters shows that observational biases are mostly restricted to the longest periods and to the most extreme mass-ratios (Fig. 1).

Compared to previous works, we find no preference for equal mass binaries (*22*). We obtain a steeper period distribution and a larger fraction of short period systems than previously thought (*9-14, 23*), resulting in a much larger fraction of systems that are affected by binary evolution.

Because star cluster dynamics and stellar evolution could have affected the multiplicity properties of only very few of the young O stars in our sample (see supporting online material §A.2), our derived distributions are a good representation of the binary properties at birth. Thus it is safe to conclude that the most common end product of massive star formation is a rather close binary. This challenges current star formation theories (*24*). However, according to recent simulations (*25-26*), accretion disk fragmentation, through gravitational instabilities, seems to naturally result in the formation of binary systems containing two massive stars with similar but not equal masses (i.e., within a factor of a few). Albeit the companions are initially formed in a wide orbit, dynamical interactions with the remnant accretion disk may significantly harden the system, providing thus a better agreement with the observations.

Intrinsic binary properties are key initial conditions for massive star evolution, i.e. evolutionary paths and final fates. Integration of our intrinsic distribution functions (see supporting online text §C and Fig. 2) implies that 71% of all stars born as O-type interact with a companion, over half of which doing so before leaving the main sequence. Such binary interactions drastically alter the evolution and final fate of the stars and appear, by far, the most frequent evolutionary channel for massive stars. Based on calculations of binary evolution in short-period systems (*6, 27-29*) we also find that 20 to 30% of all O stars will merge with their companion, and that 40 to 50% will be either stripped of their envelope or will accrete substantial mass (see supporting text §C). In summary, we find that almost three quarters of all massive stars are strongly affected by binary interaction before they explode as supernovae.

The interaction and merger rates that we computed are respectively two and three times larger than previous estimates (*6, 11, 23*). This results in a corresponding increase in the number of progenitors of key astrophysical objects which are thought to be produced by binary interaction such as close double compact objects, hydrogen-deficient CCSNe and GRBs.

We predict that 33% of O stars are stripped of their envelope before they explode as hydrogen-deficient CCSNe (Types Ib, Ic and IIb). This fraction is close to the observed fraction of hydrogen-poor supernovae, i.e. 37% of all CCSNe (*30*). Extrapolation of our findings from O stars to the 8-15 solar mass range to include all CCSN progenitors implies that hydrogen-poor

CCSNe predominantly result from mass transfer in close binaries. This rate is large enough to explain the discrepancy between the large observational fraction of Type Ib/c supernovae and the dearth of single stars stripped by stellar winds. Our results also imply that more than half of the progenitors of hydrogen-rich (Type II) supernovae are merged stars or binary mass gainers, which might explain some of the diversity of this supernova class.

Our results further indicate that a large fraction of massive main sequence stars (about 40%) is expected to be spun-up either by accretion or coalescence. In lower metallicity galaxies these stars should remain rapidly rotating and hence constitute a major channel for the production of long-duration GRBs (*31*) which are thought to accompany the death of massive stars in case their iron cores collapse to critically rotating neutron stars or black holes (*32-33*).

In conclusions, we show that only a minority of massive stars evolve undisturbed towards their supernova explosion. The effects of binarity must thus be considered in order to further our understanding of the formation and evolution of massive stars and to better interpret the integrated properties of distant star-forming galaxies (*34-35*).

**Acknowledgments:** Support for this work was provided by NASA through Hubble Fellowship grant HST-HF-51270.01-A awarded by the Space Telescope Science Institute, which is operated by the Association of Universities for Research in Astronomy, Inc., for NASA, under contract NAS 5-26555. MG acknowledges financial support of the Royal Society. This work is based on ESO observations and we acknowledge support from the ESO Paranal observatory and User Support Department. HS also acknowledge support from the SARA Computing and Networking Services. Measurements used in this work are made available at the Centre de Données astronomiques de Strasbourg (*http://cdsweb.u-strasbg.fr/*).

**Supplementary Materials:**

Materials and Methods

Figures S1-S4

Tables S1-S4

References (*36-61*)

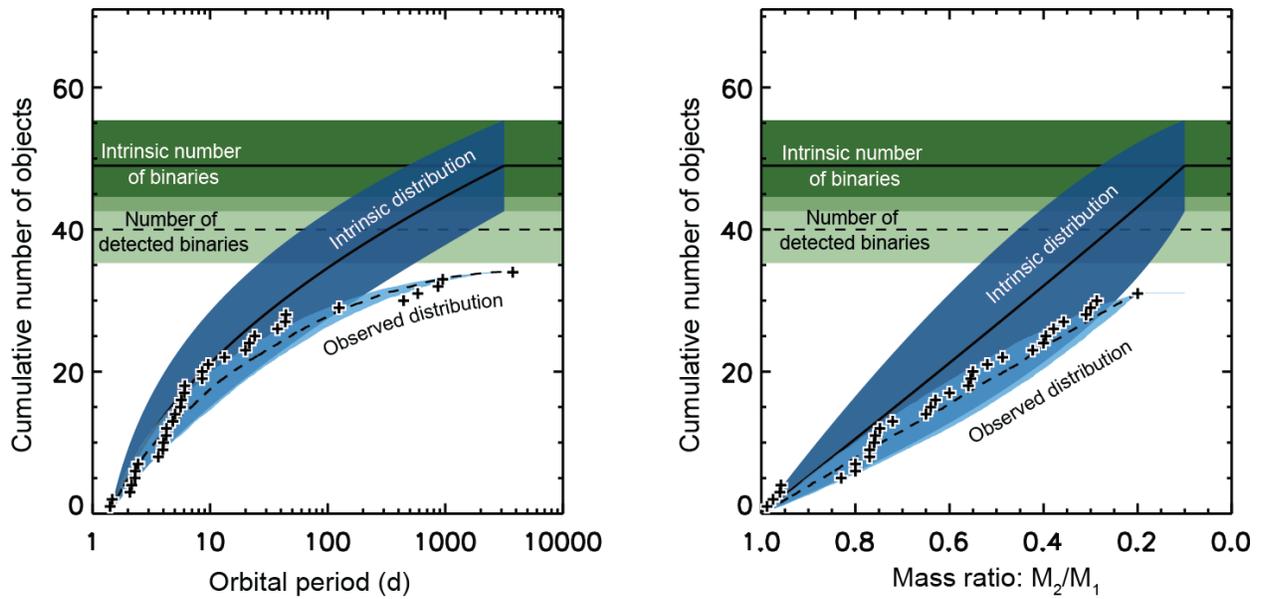

**Fig. 1**. Cumulative number distributions of logarithmic orbital periods (left panel) and of mass ratios (right panel) for our sample of 71 O-type objects, of which 40 are identified binaries. The horizontal solid line and the associated dark green area indicate the most probable intrinsic number of binaries (49 in total) and its 1σ uncertainty, corresponding to an intrinsic binary fraction $f_{bin}$ = 0.69 ± 0.09. The horizontal dashed line indicates the most probable simulated number of detected binaries: 40 ± 4, which agrees very well with the actual observed number of binaries (40 in total).

Crosses show the observed cumulative distributions for systems with known periods (34 in total) and mass-ratios (31 in total). The dashed lines indicate the best simulated observational distributions and their 1σ uncertainties. They correspond to intrinsic distributions with power law exponents π = -0.55 ± 0.22 and κ = -0.10 ± 0.58 respectively. The solid lines and associated dark blue areas indicate the most probable intrinsic number distributions and their errors. The latter were obtained from a combination of the uncertainties on the intrinsic binary fraction and on the power law exponents of the respective probability density functions.

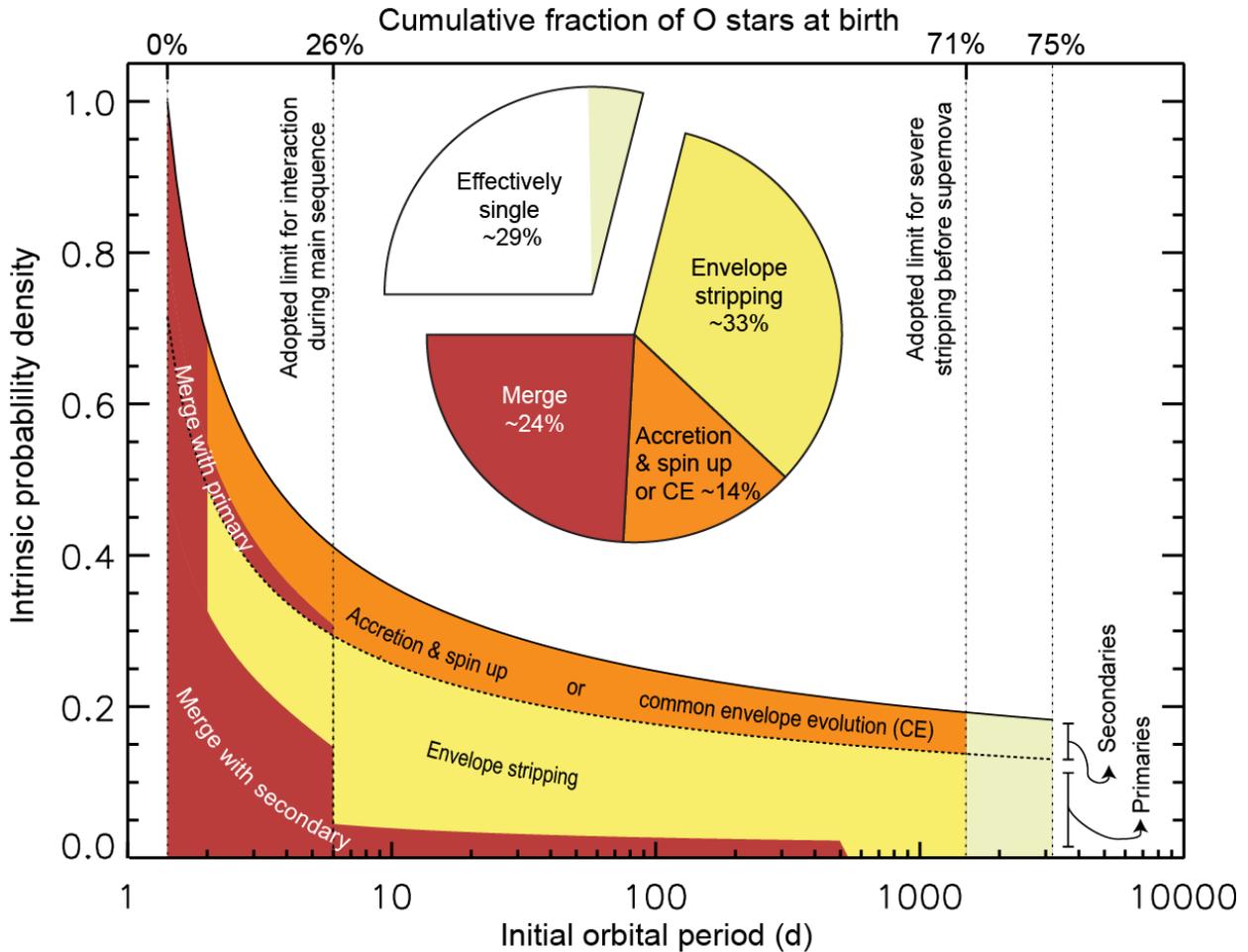

**Fig. 2.** Schematic representation of the relative importance of different binary interaction processes given our best-fit binary fraction and intrinsic distribution functions. All percentages are expressed in terms of the fraction of all stars born as O-type stars, including the single O stars and the O stars in binaries, either as the initially more massive component (the primary), or the less massive one (the secondary).

The solid curve gives the best-fit intrinsic distribution of orbital periods (corresponding to π = -0.55), which we adopted as the initial distribution. For the purpose of comparison, we normalized the ordinate value to unity at the minimum period considered. The dotted curve separates the contributions from O-type primary and secondary stars. The colored areas indicate the fractions of systems that are expected to merge (red), to experience stripping (yellow) or accretion/common envelope evolution (orange). Assumptions and uncertainties are discussed in the text and in the supporting online text §C.

The pie chart compares the fraction of stars born as O stars that are effectively single, i.e. single (white) or in wide binaries with little or no interaction effects (light green) − 29% combined − with those that experience significant binary interaction (71% combined).